\documentclass[english,twocolumn, prl, superscriptaddress, longbibliography]{revtex4-1}
\usepackage[T1]{fontenc}
\usepackage[latin9]{inputenc}
\usepackage{color}
\usepackage{amsmath}
\usepackage{graphicx}

\makeatletter
\usepackage{amsmath}
\usepackage{subfigure}
\usepackage{graphicx, mathtools}
\usepackage[colorlinks=true,linkcolor=MidnightBlue,urlcolor=MidnightBlue,citecolor=MidnightBlue,anchorcolor=MidnightBlue]{hyperref}\usepackage[dvipsnames]{xcolor}

\usepackage{chngcntr}

\makeatother

\usepackage{babel}
\begin{document}

\title{Direct verification of the fluctuation-dissipation relation in viscously
coupled oscillators}

\author{Shuvojit Paul$ $}

\affiliation{Indian Institute of Science Education and Research, Kolkata}

\author{Abhrajit Laskar}

\affiliation{The Institute of Mathematical Sciences-HBNI, CIT Campus, Taramani,
Chennai 600113, India}

\author{Rajesh Singh}

\affiliation{The Institute of Mathematical Sciences-HBNI, CIT Campus, Taramani,
Chennai 600113, India}

\author{Basudev Roy}

\affiliation{Department of Physics, Indian Institute of Technology Madras, Chennai
600036, India}

\author{R. Adhikari}
\email{rjoy@imsc.res.in}

\affiliation{The Institute of Mathematical Sciences-HBNI, CIT Campus, Taramani,
Chennai 600113, India}

\affiliation{DAMTP, Centre for Mathematical Sciences, University of Cambridge,
Wilberforce Road, Cambridge CB3 0WA, UK}

\author{Ayan Banerjee}
\email{ayan@iiserkol.ac.in}

\affiliation{Indian Institute of Science Education and Research, Kolkata}

\date{\today}
\begin{abstract}
The fluctuation-dissipation relation, a central result in non-equilibrium
statistical physics, relates equilibrium fluctuations in a system
to its linear response to external forces. Here we provide a direct
experimental verification of this relation for viscously coupled oscillators,
as realized by a pair of optically trapped colloidal particles. A
theoretical analysis, in which interactions mediated by slow viscous
flow are represented by non-local friction tensors, matches experimental
results and reveals a frequency maximum in the amplitude of the mutual
response which is a sensitive function of the trap stiffnesses and
the friction tensors. This allows for its location and width to be
tuned and suggests the utility of the trap setup for accurate two-point
microrheology.
\end{abstract}
\maketitle
The relation between the generalized susceptibility and equilibrium
fluctuations of the generalized forces, first obtained for a linear
resistive circuit by Nyquist \cite{nyquist1928thermal} and then proved
for any general linear dissipative system by Callen and Welton \cite{callen1951},
is a central result in non-equilibrium statistical physics. The relation
can be used to infer the intrinsic fluctuations of a system from measurements
of its response to external perturbations or, perhaps more startlingly,
to predict its response to external perturbations from the character
of its intrinsic fluctuations \cite{kubo1966fluctuation}. The fluctuation-dissipation
relation is the point of departure for several areas of current research
including fluctuation relations \cite{seifert2012stochastic}, relaxation
in glasses \cite{berthier2011theoretical}, and response and correlations
in active \cite{fodor2016far} and driven systems \cite{mason95,levine2000}. 

The first experimental\emph{ }verification of the relation between
fluctuation and dissipation was due to Johnson \cite{johnson1928thermal},
whose investigation of the ``thermal agitation of electricity in
conductors'' provided the motivation for Nyquist's theoretical work
\cite{nyquist1928thermal}. Though the relation has been verified
since in systems with conservative couplings, a direct verification
in a system where the coupling is entirely dissipative is, to the
best of our knowledge, not available. Colloidal particles in a viscous
fluid interact through velocity-dependent many-body hydrodynamic forces
whose strength, away from boundaries, is inversely proportional to
the distance between the particles. The range of these dissipative
forces can be made much greater than that of conservative forces such
as the DLVO interaction \cite{derjaguin1941theory,verwey1948}. Therefore,
it is possible to engineer a situation where the dominant coupling
between colloidal particles is the viscous hydrodynamic force and
all other interactions are negligibly small. Such systems, then, are
ideal for testing the fluctuation-dissipation relation when couplings
are purely dissipative. 

In this Letter, we present a direct verification of the fluctuation-dissipation
relation for a pair of optically trapped colloidal particles in water.
We measure the equilibrium fluctuations of the distance between the
particles and the response of one particle to the sinusoidal motion
of another particle. Transforming both correlations and responses
to the frequency domain, we verify the fluctuation-dissipation relation
over a range of frequencies spanning two orders of magnitude. Remarkably,
the response function has a peak in frequency, reminiscent of a resonance,
though the system of oscillators is entirely overdamped. A theoretical
analysis, assuming slow viscous flow of the ambient water, is in excellent
agreement with the experiments. The analysis reveals that the location
and width of the resonant peak can be tuned by altering the viscosity,
the separation between the particles, the trap stiffnesses, and the
colloidal diameters. It provides the inverse relations necessary for
using the trap setup for accurate two-point microrheology. We now
present details of our experiment and its analysis.
\begin{figure}
\centering \includegraphics[width=0.485\textwidth]{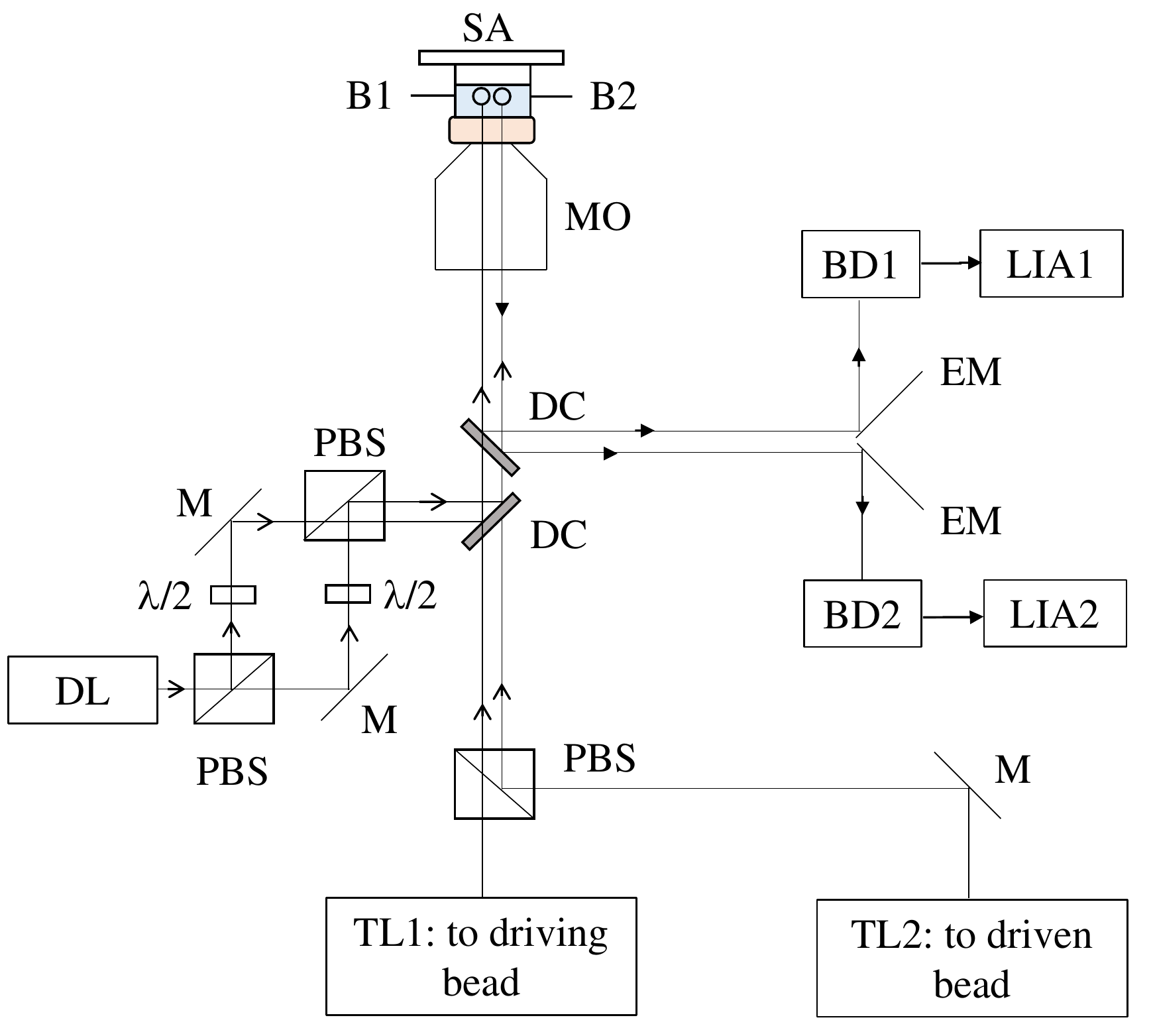} \caption{A schematic diagram of the experimental setup. TL1: trapping laser
for driving particle B1, TL2: trapping laser for driven particle B2,
DL: detection laser, PBS: polarizing beam splitter cube, $\frac{\lambda}{2}$:
half wave plate, DC: dichroic mirror, MO: microscope objective, CS:
cover slip, BD1 and BD2: balanced detection systems based on Thorlabs
photodiodes PD-EC2, M: mirror, EM: edge mirror, LIA1 and LIA2: lock-in
amplifiers for B1 and B2, respectively.}
\label{Setup} 
\end{figure}

\textit{Experiment:} The details of the experimental setup towards
validation of the fluctuation-dissipation theorem are provided in
Supplementary Information - here we provide a brief description. Thus,
we set up a dual-beam optical tweezers (Fig.\ref{Setup}) by focusing
two orthogonally polarized beams of wavelength $\lambda=1064$ nm
generated independently from two diode lasers using a high NA immersion-oil
microscope objective (Zeiss PlanApo,$100\times1.4$). One of the lasers
is modulated using an AOM located conjugate to the back-focal plane
of the microscope objective, and a long optical path after the AOM
ensures that a minimal beam deflection is enough to modulate one of
the trapped beams, so that the intensity in the first order remains
constant to around 2\%. The modulated and unmodulated beams are independently
coupled into the trapping microscope using mirrors and a polarizing
beam splitter, while detection is performed using a separate laser
at 671 nm generating two detection beams also orthogonally polarized
and superposed on the respective trapping beams using dichroic beam
splitters. The two trapped beads are imaged and their displacements
measured by back-focal- plane-interferometry, with the imaging white
light and detection beams also separated at the output by dichroic
beam splitters, which along with the orthogonal polarization scheme
ensures that cross-talk in the detection beams is absent. A very low
volume fraction sample ($\phi\approx0.01$) is prepared with 3 $\mu$m
diameter polystyrene latex beads in 1 M NaCl-water solution for avoiding
surface charges. We trap two spherical polystyrene beads (Sigma LB-30)
of mean size 3 $\mu$m each, in two calibrated optical traps which
are separated by a distance $4\pm0.1\:\mu$m, so that the surface-surface
distance of the trapped beads is $1\pm0.2\:\mu$m ($0.67a$, $a$
being the particle radius) and the distance from the cover slip surface
is 30 $\mu$m ($20a$, so as to overrule wall effects). From the literature
\cite{Stilgoe11}, the particle separation is still large enough to
avoid effects due to optical binding and surface charges. In order
to ensure that the trapping and detection beams are not influencing
each other, we measure the Brownian motion of a trapped particle when
the trapping and detection beams for the other trap is switched on
(in the absence of a particle), and check that there are no changes
in the measured trap stiffness. One of the traps is sinusoidally modulated
(amplitude around $0.2a$) and the phase and amplitude response of
both the driving and driven particles with reference to the sinusoidal
drive are measured by lock-in detection (Stanford Research, SR830).
To get large signal to noise, we use balanced detection systems BD1
and BD2, for the driving and driven particles, respectively. The voltage-amplitude
calibration of our detection system reveals that we can resolve motion
of around 5 nm with an SNR of 2.
\begin{figure*}
\centering \includegraphics[width=0.98\textwidth]{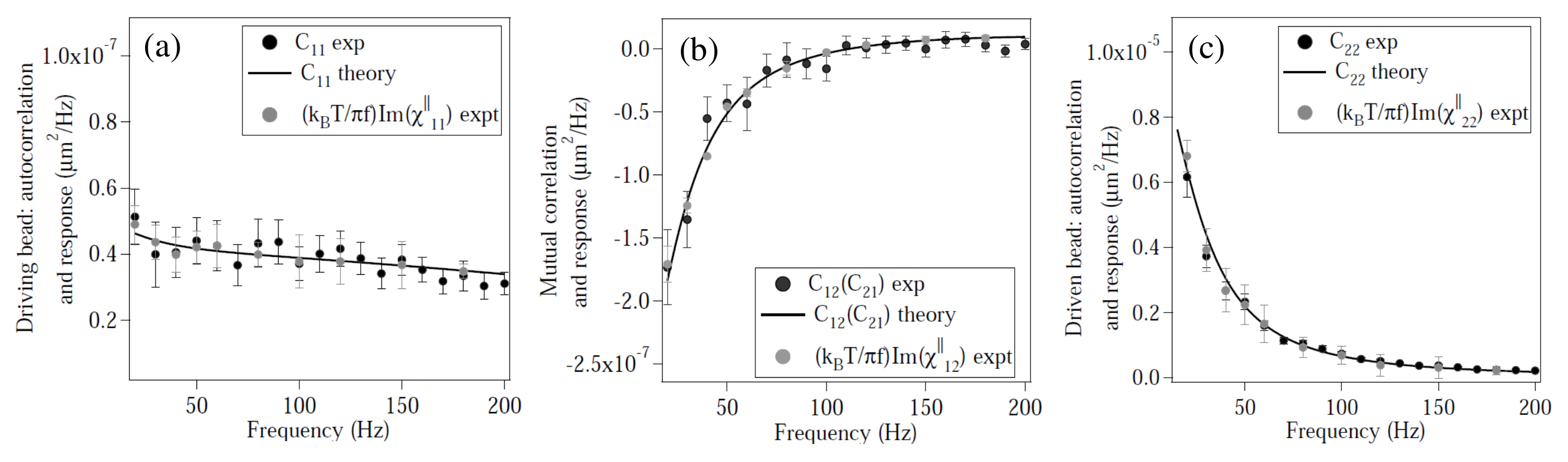} \caption{Verification of the fluctuation-dissipation relation $C_{ij}=(k_{B}T/\pi f)\,\text{Im}(\chi_{ij}^{\parallel})$
for a pair of viscously coupled colloidal particles in optical traps.
The first and third panels compare the self-response and the position
auto-correlation of, respectively, the driving and driven colloid,
while the second panel compares their mutual-response and cross-correlation.
Theoretically computed correlation functions, assuming over-damped
motion of the colloids and slow viscous flow in the fluid, are shown
as solid lines. }
\label{fdt} 
\end{figure*}

Each of the optical traps are calibrated using equipartition and power
spectrum methods considering the particle temperature to be same as
the room temperature. We verify that each of the potentials is harmonic
in nature from the histogram of the Brownian motion which is satisfactorily
Gaussian (Fig.\ref{fig:SI-1} in Supplementary Information), even
when both trapping beams are on. The sampling frequency is 2 kHz,
while we performed data blocking at the level of 100 points in order
to ensure good Lorentzian fits \cite{berg2004power} for trap calibration.
We maintain a considerably higher stiffness for the particle in the
modulated trap so that it is not affected by the back-flow due to
the driven particle. The low stiffness of the driven trap ensures
that it has a maximal response to the drive. Thus, for validation
of the fluctuation-dissipation theorem, the stiffness of the modulated
bead (B1) was 69.6 $\mu N/m$, while that of the driven is 4.8 $\mu N/m$.
Note that, to observe a clear amplitude resonance, a lower ratio of
trap stiffness is required, as we demonstrate later. The verification
of the fluctuation-dissipation theorem is shown in Fig.\ref{fdt}.
It is understandable that while the fluctuation-dissipation theorem
is in the form a simple equation for a single particle, for two particles
the equations would be represented in the form of a matrix, which
we discuss in more detail later. This is what we demonstrate in Fig.\ref{fdt}(a),
(b), and (c), where the auto and cross-correlations for both particles
are matched with the corresponding response functions. The auto-correlation
function of B1 is shown in Fig.\ref{fdt}(a), while that of B2 is
shown in Fig.\ref{fdt}(b). The corresponding response functions ($\mathit{\chi_{11}^{||},\:\chi{}_{22}^{||}}$)
are obtained by measuring the amplitude and phase of the individual
particles when they are themselves driven. Fig.\ref{fdt}(c) shows
the cross-correlation function which is again compared with the corresponding
response function $\mathit{\chi_{12}^{||}}$. This is obtained by
measuring the amplitude and phase of B2 when B1 is driven. Note that
we are not able to measure $\mathit{\chi_{21}^{||}}$which is the
response of B1 when B2 is driven since the much larger stiffness of
B1 renders the amplitude of the response extremely small so that it
is beyond our detection sensitivity. For the response measurements,
each data point is the average of ten separate measurements at each
frequency. It is clear from the figures that we obtain a good match
between fluctuation and response - which essentially validates the
fluctuation-dissipation relations for a pair of colloidal particles
coupled by hydrodynamic interactions. Note that for consistency check,
we also plot the cross-correlation function in the time domain (Fig.\ref{fig:SI-4}
in Supplementary Information) and obtain qualitatively similar data
as reported in Ref. \cite{Meiners99}.

\textit{Theory:} The Langevin equations describing the stochastic
trajectories of the colloids are \cite{gardiner1985handbook}
\begin{align}
m_{i}\dot{\boldsymbol{v}_{i}} & +\boldsymbol{\gamma}_{ij}\cdot\boldsymbol{v}_{j}+{\bf \boldsymbol{\nabla}}_{i}U=\boldsymbol{{\bf \xi}}_{i},\label{eq1}
\end{align}
where $i,j=1,2$ refer to the driving and driven colloid, $m_{i}$
are their masses, $\boldsymbol{v}_{i}$ are their velocities, $\boldsymbol{\gamma}_{ij}$
are the second-rank friction tensors encoding the velocity-dependent
dissipative forces mediated by the fluid, $U=U_{1}+U_{2}$ is the
total potential of the conservative forces, and $\boldsymbol{\xi}_{i}$,
the Langevin noises, are zero-mean Gaussian random variables whose
variance is provided by the fluctuation-dissipation relation $\langle\boldsymbol{\xi}_{i}(t)\boldsymbol{\xi}_{j}(t^{\prime})\rangle=2k_{B}T\boldsymbol{\gamma}_{ij}\delta(t-t^{\prime})$.
The bold-face notation, with Cartesian indices suppressed, is used
for both vectors and tensors.

In the limit of slow viscous flow in the fluid, the friction tensors
can be calculated from the Stokes equation using a variety of methods
\cite{ladd1988,mazur1982,cichocki1994friction,singh2016crystallization,singh2016traction}.
To leading order the result is
\begin{equation}
\boldsymbol{\gamma}_{ij}=\delta_{ij}\boldsymbol{I}\gamma_{i}-(1-\delta_{ij})\gamma_{i}\gamma_{j}\mathcal{F}_{i}\mathcal{F}_{j}\boldsymbol{G}(\boldsymbol{r}_{i},\boldsymbol{r}_{j}),
\end{equation}
where $\gamma_{i}=6\pi\eta a_{i}$ are the self-frictions, $\boldsymbol{G}$
is a Green's function of the Stokes equation \cite{pozrikidis1992},
$\boldsymbol{r}_{i}$ are the centers of the colloids and $\mathcal{F}_{i}=1+\frac{a_{i}^{2}}{6}\nabla_{i}^{2}$
are the Faxén corrections that account for the finite radius, $a_{i}$,
of the colloids. We emphasize that this expression is not limited
to the translationally invariant Green's function of unbounded flow,
$8\pi\eta\,\boldsymbol{G}(\boldsymbol{r})=(\nabla^{2}\boldsymbol{I}-\boldsymbol{\nabla\nabla})r$,
but holds generally for any Green's function and is both symmetric
and positive-definite \cite{singh2016crystallization,singh2016traction}.
The mutual friction tensors decay inversely with distance in an unbounded
fluid and more rapidly in the proximity of boundaries. The assumption
of slow viscous flow is valid at frequencies $\omega\tau_{\nu}\ll1$
where $\tau_{\nu}=\rho L^{2}/\eta$ is the vorticity diffusion time
scale \cite{kim2005}. 

The harmonic optical potentials are given by $U_{i}(t)=\frac{1}{2}k_{i}|{\bf \boldsymbol{r}}_{i}-{\bf \boldsymbol{r}}_{i}^{0}|^{2}$
where $\boldsymbol{r}_{i}^{0}$ are the centers and $k_{i}$ are the
stiffnesses of the optical traps. Note the absence of conservative
mutual couplings. The system remains in equilibrium when the trap
centers are stationary but is driven into non-equilibrium when they
are modulated in time as $\boldsymbol{r}_{i}^{0}(t)$. For small modulations
the response is linear. 

For modulation frequencies $\omega\ll\gamma_{i}/m_{i}$ the velocities
can be adiabatically eliminated from the inertial Langevin equations
to yield inertialess Langevin equations for the positions \cite{gardiner1984adiabatic}.
The multiplicative noises in the resulting equations have clear interpretations
within the adiabatic elimination procedure; there is no Itô-Stratonovich
dilemma \cite{van1981ito,gardiner1985handbook,van1992stochastic,klimontovich1990ito,klimontovich1994nonlinear}.
Both correlation and response functions can be calculated in this
limit. Linearizing about the mean separation between the trap centers
and decomposing the motion into components parallel and perpendicular
to the separation vector, the result for the parallel response function
is
\begin{gather}
\text{Im}\left[\chi_{ij}^{||}(\omega)\right]=\frac{\omega M_{ij}}{(\text{det}A-\omega^{2})^{2}+(\omega\begin{tabular}{c}
 tr\end{tabular}A)^{2}},
\end{gather}
where $A_{ij}=\mu_{ij}^{||}k_{j}$ is a ``response'' matrix, the
mobility matrix $\mu_{ij}^{||}$ is the inverse of the friction matrix
and
\[
M_{ij}=\left(\begin{array}{cc}
\begin{tabular}{c}
 \ensuremath{\tfrac{k_{2}}{k_{1}}\mu_{22}^{\parallel}}\end{tabular}\text{det}A+\mu_{11}^{\parallel}\omega^{2} & -\mu_{12}^{\parallel}(\text{det}A-\omega^{2})\\
\, & \,\\
-\mu_{21}^{\parallel}(\text{det}A-\omega^{2}) & \begin{tabular}{c}
 \ensuremath{\tfrac{k_{1}}{k_{2}}\mu_{11}^{\parallel}}\end{tabular}\text{det}A+\mu_{22}^{\parallel}\omega^{2}
\end{array}\right).
\]
The magnitude of the response of the driven bead to the driving bead
is maximum at the ``resonance'' frequency
\begin{equation}
\omega_{res}=\sqrt{\text{det}A}=\sqrt{\mu_{11}^{\parallel}\mu_{22}^{\parallel}k_{1}k_{2}\left(1-\frac{\mu_{12}^{\parallel}\mu_{21}^{\parallel}}{\mu_{11}^{\parallel}\mu_{22}^{\parallel}}\right)}.\label{eq4}
\end{equation}
A simple analysis of the system with the two particles executing Brownian
motion in the absence of the external drive leads us to write down
the auto and cross-correlation functions ($C_{ij}$), so that by comparing
with the response functions $\text{Im}(\chi_{ij}^{\parallel})$, we
have $C_{ij}=(k_{B}T/\pi f)\,\text{Im}(\chi_{ij}^{\parallel})$ which
is the well known fluctuation-dissipation relationship \cite{kubo1966fluctuation}.
\begin{figure}
\centering \includegraphics[width=0.48\textwidth]{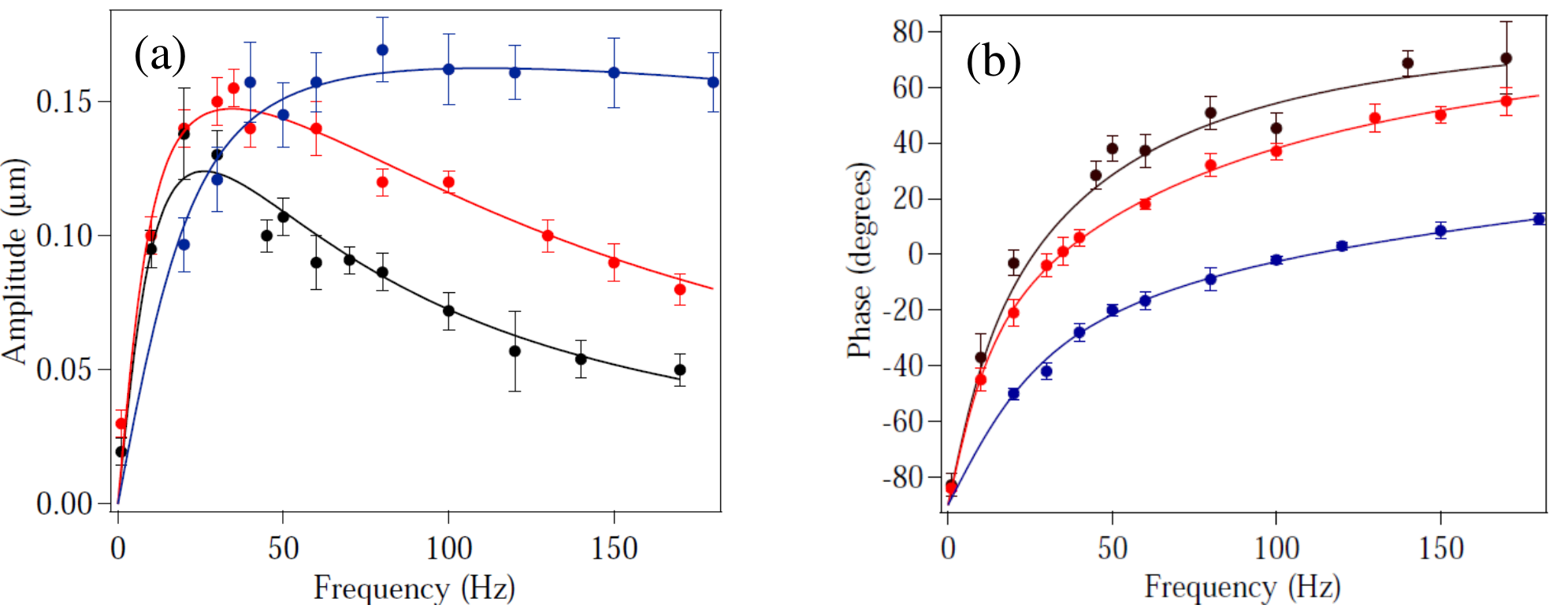} \caption{Amplitude and phase response for driven (B2) bead for different trap
stiffness ratios with the inter-particle separation $0.67a$. (a)
and (b) demonstrate amplitude and phase responses (with respect to
driving frequency) of B2, for trap stiffness ratios of 2.5:1 (black),
5.7:1 (red), and 14.5:1 (blue). The resonance frequency in (a) is
20 Hz (black), 35 Hz (red), and 111 Hz (blue). The solid spheres denote
experimental data points while the solid lines are corresponding theoretical
fits. }
\label{response}
\end{figure}
 This is indeed what we validate in Fig.\ref{fdt}(a)-(c). 

We now focus on a particularly interesting facet of our problem, namely
the amplitude and phase response of B2 under the influence of the
driven particle B1. We study this experimentally for three different
trap stiffness ratios of B1 and B2, the results of which are shown
in Fig.\ref{response}(a) and (b). Note that we fit each graph with
the calculated values of the responses for the experimental parameters
used, and obtain very good fits. The amplitude and phase response
of B1 (Supplementary information) to the drive frequency is expected,
with the amplitude decaying with increasing frequency, and the phase
being in sync with the drive at low frequencies and gradually lagging
behind as the frequency is increased. However, the amplitude response
of B2 is rather interesting, and shows a clear resonance response
at a certain frequency, the value of which increases as the stiffness
ratio of the traps is increased - it being dependent on the product
of the stiffnesses as is clear from Eq.\ref{eq4}. Thus, we have a
resonance frequency of around 111 Hz (blue solid spheres in Fig.\ref{response}(a))
with $\mathit{k_{1}:k_{2}}=$14.5:1, a frequency of around 33 Hz with
a ratio of 5.7:1 (red solid spheres), and a frequency of around 20
Hz with a ratio of 2.5:1 (black solid spheres). In fact, it is as
if the entrained fluid has minimum impedance around this frequency,
so that there is maximum energy transfer between the driving and the
driven beads. The amplitude of the resonance has an inverse dependence
on the particle separation, so that with our current detection sensitivity,
we do not observe the resonance effects beyond a surface-surface separation
greater than $3a$. However, even this increased distance is also
smaller than that used in earlier experiments, which possibly explains
the fact that this phenomenon has not been reported earlier. The width
of the resonance ($\mathit{Q}$ factor) is also dependent on the stiffness
ratio, and increases as the latter is reduced. For a given medium,
the resonance can thus be tuned by changing the stiffness ratios (as
well as the inter-particle separation and particle diameters). Interestingly,
it is obvious that the value of the $\mathit{Q}$-factor as well as
the resonance frequency also depends on the damping, and can be modified
by changing the viscosity of the solution. This property promises
the measurement of this frequency shift as an accurate two-point micro-rheology
probe of local viscosity of a fluid. Finally, the phase response in
\ref{response}(b) is easily explained: B2 lags 90 degrees in phase
with respect to the drive at very low frequencies with the lag reducing
until the drive and driven are in phase at resonance, after which
the driven bead leads in phase, and asymptotically approaches 90 degrees
at high frequencies. The rate of approach is also determined by the
stiffness ratio, and is rather slow at large stiffness ratios. Indeed,
this is exactly similar to the relationship between velocity and driving
force for a forced damped harmonic oscillator, and arises due to the
fact that the oscillators are dissipatively coupled. 

In conclusion, we perform a direct experimental verification of the
fluctuation-dissipation relation in a system consisting of two colloidal
particles confined in a viscous medium (water) in very close proximity
(surface-surface separation less than the particle radius) using separate
optical tweezers. Our results provide a confirmation of the validity
of the fluctuation-dissipation relation in the presence of long-ranged
dissipative forces that are the only source of coupling of, otherwise,
independent degrees of freedom. Surprisingly, we identify a resonance
in the response in a system which is overdamped and suggest its use
in accurate two-point microrheology. The present experiment can be
extended in several directions: measurements at higher frequencies
can uncover the effects of retarded hydrodynamic interactions and
the role of particle inertia while holographic traps can be used to
test the fluctuation-dissipation relation in the presence of many-body
hydrodynamic interactions. Some of these will be presented in forthcoming
work. 

This work was supported by the Indian Institute of Science Education
and Research, Kolkata, an autonomous research and teaching institute
funded by the Ministry of Human Resource Development, Govt. of India.
We acknowledge computing resources on the Annapurna cluster provided
by The Institute of Mathematical Sciences. \vspace*{0.16cm}

\appendix
\begin{center}
\textbf{{\Large{Supplemental information}}}
\end{center}
\counterwithin{figure}{section} 

\section{Experiment}

\begin{figure}
\centering \includegraphics[width=0.49\textwidth]{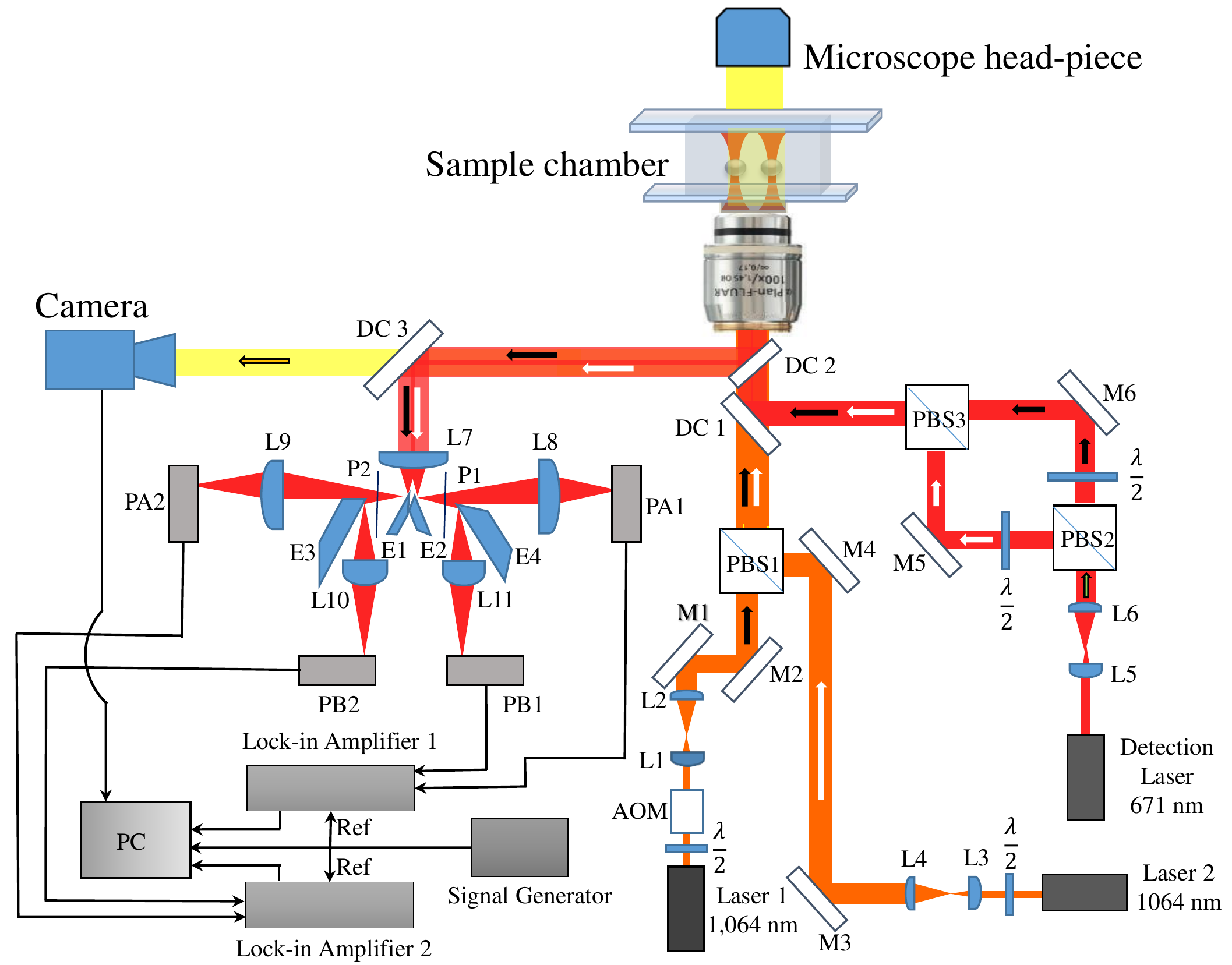} 

\caption{A detailed schematic of the setup is shown. Key:$\frac{\lambda}{2}$:
half-wave plate, L: lens, M: mirror, AOM: accousto optical modulator,
PBS: polarising beam splitter, DC: dicroic mirror respectively, E:
edge mirror, P: polariser, PA, PB: photodiodes. }

\label{fig:SI-1} 
\end{figure}

We set up a dual-beam optical tweezers (Fig.\ref{fig:SI-1}) by focusing
two orthogonally polarized beams of wavelength $\lambda=1064$ nm
generated independently from two diode lasers using a high NA immersion-oil
microscope objective (Zeiss PlanApo,$100\times1.4$). An AOM, located
conjugate to the back-focal plane of the objective using the telescopic
lens pair L1-L2 (see Fig.\ref{fig:SI-1}), is used for modulating
one of the traps. A long optical path after the AOM ensures that a
minimal beam deflection is enough to modulate one of the trapped beams,
so that the intensity in the first order remains constant to around
2\%. The modulated and unmodulated beams are independently steered
using mirror pairs M1, M2 and M3, M4, respectively, and coupled into
a polarizing beam splitter (PBS1). For detection, we use a separate
laser of wavelength $\lambda=671$ nm, that is again divided into
two beams of orthogonal polarization by PBS2 and coupled into PBS3.
We then use a dichroic (DC1) to overlap the two pairs of trapping
and detection beams into the optical tweezers microscope (Zeiss Axiovert.A1).
The two trapped beads are imaged and their displacements measured
by back-focal- plane-interferometry, while the white light for imaging
and the detection laser beams are separated by dichroics DC2 and DC3,
respectively. A very low volume fraction sample ($\phi\approx0.01$)
is prepared with 3 $\mu$m diameter polystyrene latex beads in 1 M
NaCl-water solution for avoiding surface charges. A single droplet
of about 20 $\mu$l volume of the sample is introduced in a sample
chamber made out of a standard 10 mm square cover slip attached by
double-sided sticky tape to a microscope slide. We trap two spherical
polystyrene beads (Sigma LB-30) of mean size 3 $\mu$m each, in two
calibrated optical traps which are separated by a distance $4\pm0.1\:\mu$m,
so that the surface-surface distance of the trapped beads is $1\pm0.2\:\mu$m,
and the distance from the cover slip surface is 30 $\mu$m. From the
literature, this distance is still large enough to avoid optical cross
talk and effects due to surface charges \cite{Stilgoe11}. In order
to ensure that the trapping beams do not influence each other, we
measure the Brownian motion of one when the other is switched on (in
the absence of a particle), and check that there are no changes in
the Brownian motion. One of the traps is sinusoidally modulated and
the phase and amplitude response of both the driving and driven particles
with reference to the sinusoidal drive are measured by lock-in detection
(Stanford Research, SR830). To get large signal to noise, we use balanced
detection using photodiode pairs PA1, PB1 and PA2, PB2, for the driving
and driven particles, respectively. The two beams for balanced detection
are prepared by edge mirrors E1, E3 (E2, E4) for the driving (driven)
particle, respectively. Polarizers P1 (P2) are aligned in such a way
so as select the desired polarization component of the detection beams
that are prepared, as mentioned above, in orthogonal polarization
states for the driving (driven) particle. Thus, we use a combination
of orthogonal polarization and dichroic beam splitters to separate
out the detection beams for the driving and driven particles, respectively.
The voltage-amplitude calibration of our detection system reveals
that we can resolve motion of around 5 nm with an SNR of 2.

\begin{figure}
\centering \includegraphics[width=0.45\textwidth]{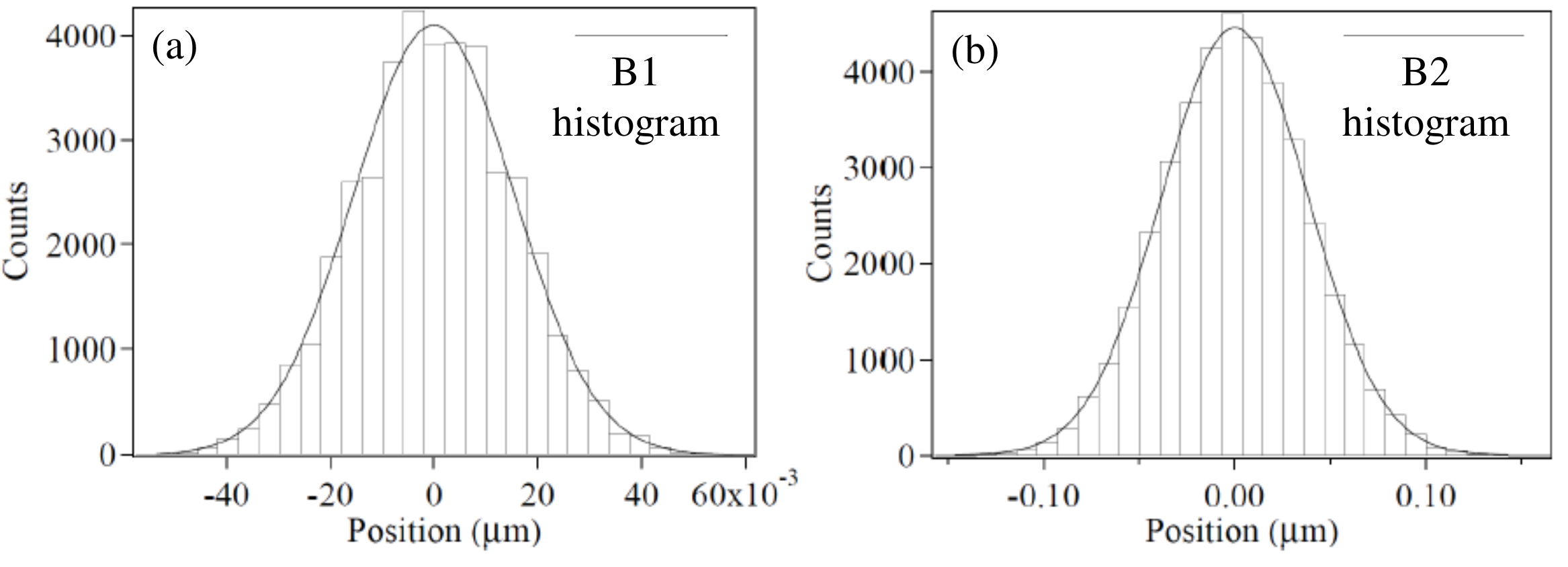} \caption{Position histograms of (a) B1 (driving particle) and (b) B2 (driven
particle). The solid black lines are corresponding Gaussian fits which
show that the potentials are harmonic in nature.}
\label{fig:SI-2} 
\end{figure}

\begin{figure}
\includegraphics[width=0.46\textwidth]{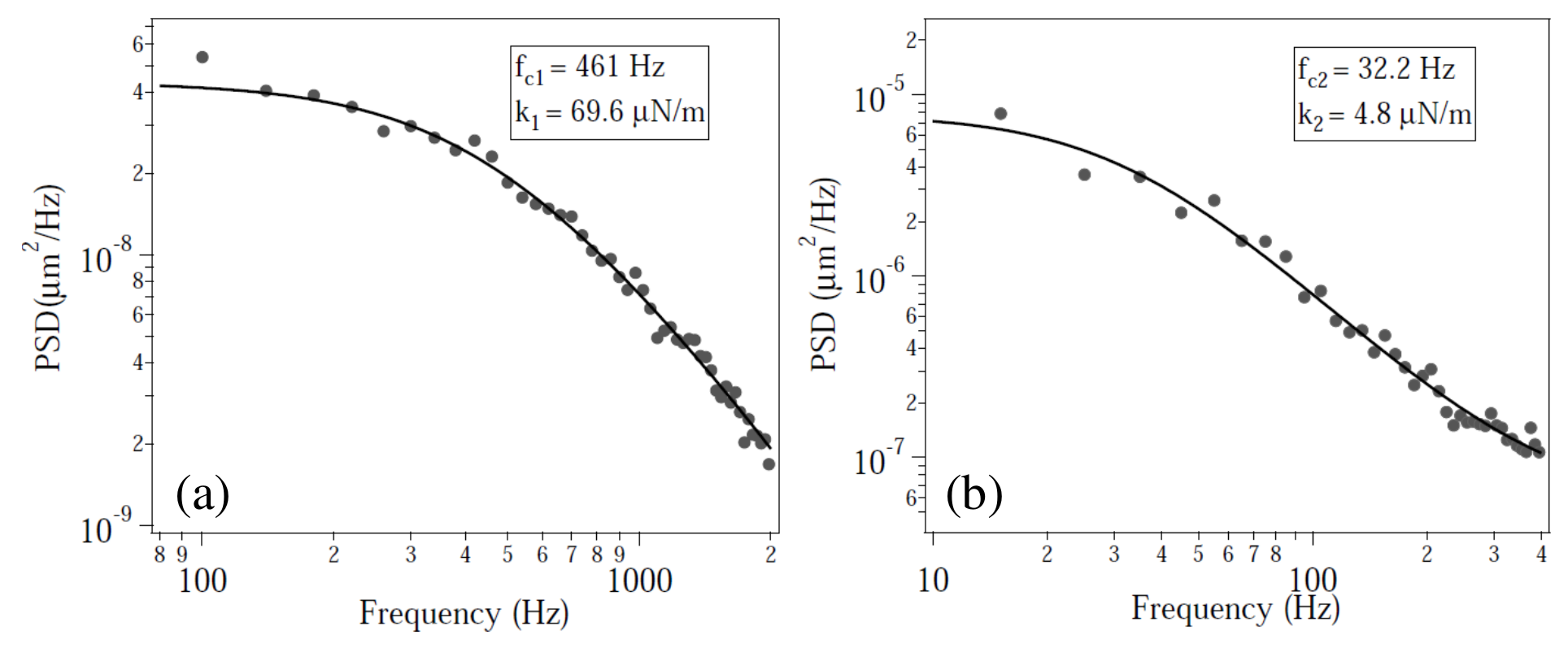}\caption{Calibration of traps for B1 and B2. Experimentally measured data points
are shown in filled gray circles, whereas the Lorentzian fit is denoted
by the solid black line. (a) PSD for B1 which has a corner frequency
$\mathit{f_{c1}=\mathrm{461}}$Hz and stiffness $\mathit{k_{1}=\mathrm{69.6}\:\mu}$N/m.
(b) PSD for B2 which has a corner frequency $\mathit{f_{c1}=\mathrm{32.2}}$Hz
and stiffness $\mathit{k_{1}=\mathrm{4.8}\:\mu}$N/m.\label{fig:SI-3}}
\end{figure}

\begin{figure}
\includegraphics[width=0.3\textwidth]{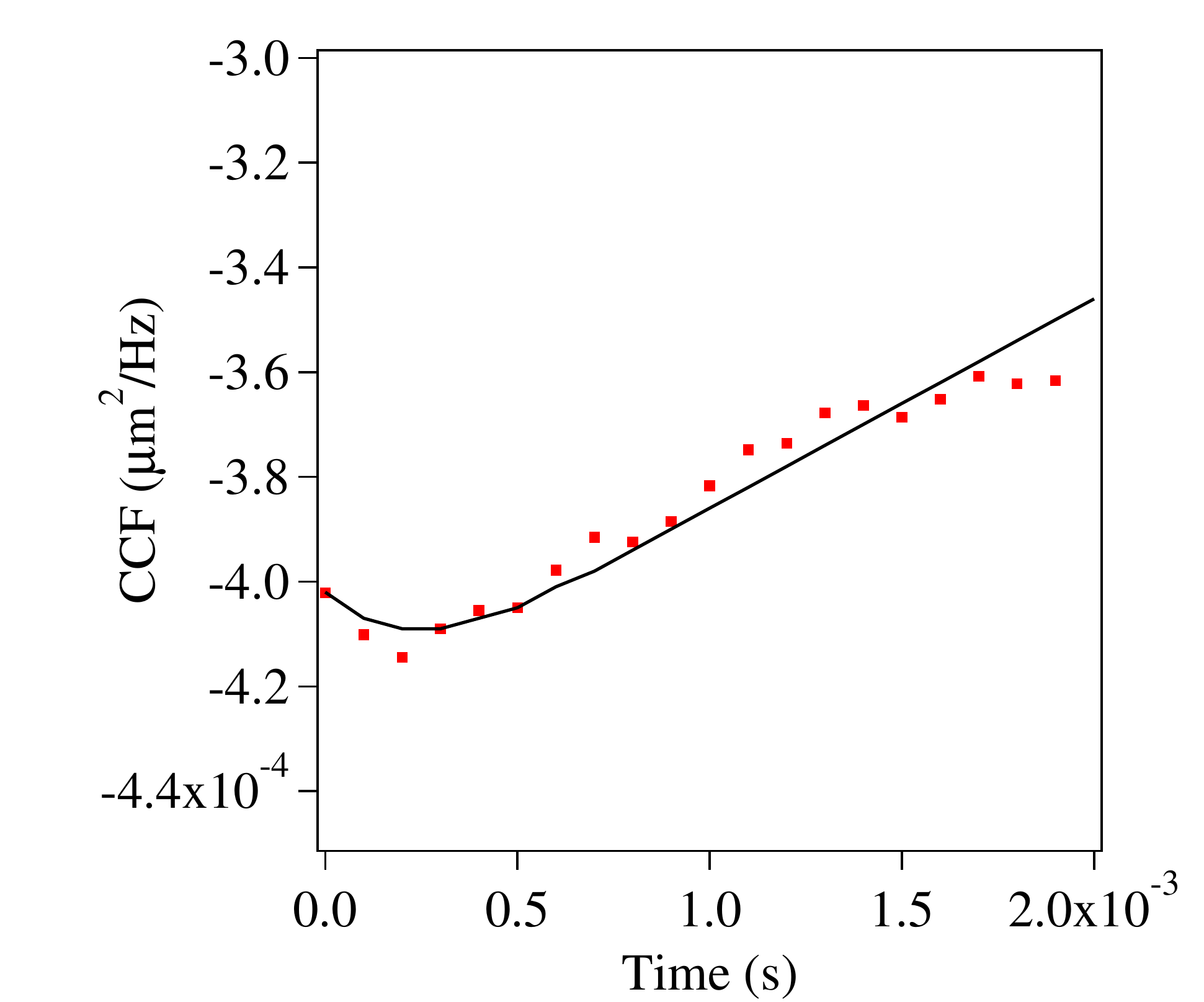}

\caption{Position cross-correlation in time domain. The filled red squares
are experimentally measured points, while the solid line is the theoretically
calculated cross-correlation.\label{fig:SI-4}}
\end{figure}

\begin{figure}
\includegraphics[width=0.46\textwidth]{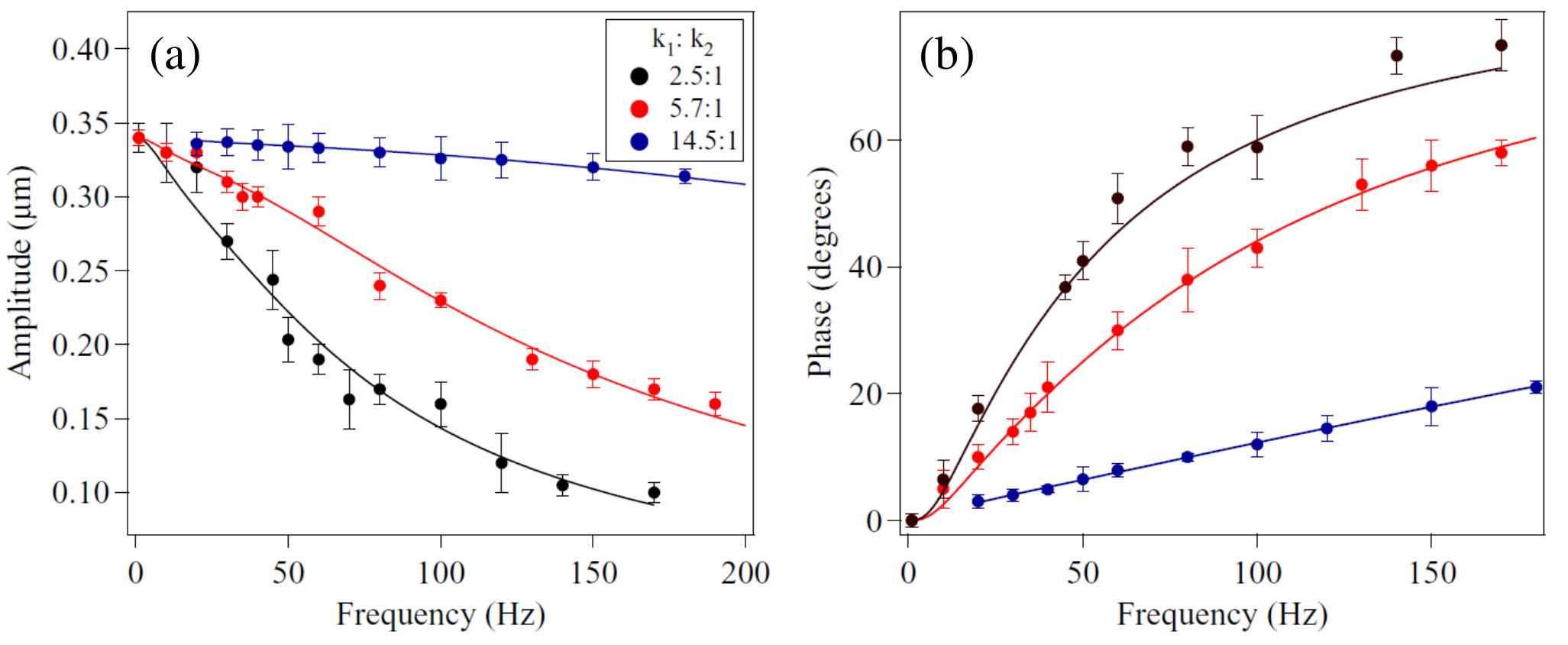}

\caption{Amplitude (a) and phase (b) response of driving particle B1 as a function
of drive frequency.\label{fig:SI-5}}
\end{figure}

Fig.\ref{fig:SI-2} (a) and (b) shows the histogram of position coordinate
data that we acquire for the Brownian motion of driving particle B1
and driven particle B2, respectively. As is clear, the data are normally
distributed in both traps and fit very well to Gaussians (shown in
bold lines). To calibrate the traps and determine the trap stiffnesses,
we measure the power spectral density (PSD) of the Brownian motion
of each particle in the absence of the other. The results are shown
in Fig.\ref{fig:SI-3}(a) and (b). Each PSD is obtained by data blocking
100 points in the manner described in Ref. \cite{berg2004power}.
The Lorentzian fits to the data are good, and we obtain corner frequencies
$\mathit{f_{c1}}=$461 Hz and $\mathit{f_{c2}}=$32.2 Hz for particles
B1 and B2, which yield stiffnesses of $\mathit{k_{1}=}$69.2 $\mu N/m$
and $\mathit{k_{2}}$= 4.8 $\mu N/m$, respectively. For the two particle
correlation experiments, as a consistency check, we determine the
position cross-correlation function in time domain for B1 and B2 as
shown in Fig.\ref{fig:SI-4}. The data fits well to Eq.5 in Ref. \cite{Meiners99},
with the constant parameters appropriately calculated for our case.
Finally, we demonstrate the amplitude and phase response of the driving
particle B1 as a function of the driving frequency in Fig.\ref{fig:SI-5}(a)
and (b), respectively. As expected, the amplitude decays with increasing
frequency, while the phase is in sync with the drive at low frequencies
and gradually lags behind as the frequency is increased.

\section{Theory}

We outline below key steps in deriving the response and correlations
functions on the Smoluchowski time scale, \emph{i.e.} the over-damped
limit, starting from Langevin equations
\begin{align}
m_{i}\dot{\boldsymbol{v}_{i}} & +\boldsymbol{\gamma}_{ij}\cdot\boldsymbol{v}_{j}+{\bf \boldsymbol{\nabla}}_{i}U=\boldsymbol{{\bf \xi}}_{i}\label{eq1-1}
\end{align}
presented and explained in the main text. 

(\emph{i}) \emph{adiabatic elimination of momentum}: in the first
step, the momenta $m_{i}\boldsymbol{v}_{i}$ are adiabatically eliminated
from the Langevin equations to obtain a contracted description in
terms of the positions alone \cite{gardiner1984adiabatic,gardiner1985handbook}.
This equation is valid on time scales $t\gg m_{i}/\gamma_{i}$. The
heuristic of setting $m_{i}\boldsymbol{v}_{i}$ to zero in the Langevin
equations yields the same result as the more systematic adiabatic
elimination procedure, provided the multiplicative noise is interpreted
correctly and the so-called ``spurious'' drift is included in the
equation for the position increment \cite{van1981ito,van1992stochastic}.
With these caveats, the resulting over-damped Langevin equations are
\begin{equation}
\boldsymbol{\gamma}_{ij}\cdot\boldsymbol{\dot{r}}_{j}+{\bf \boldsymbol{\nabla}}_{i}U=\boldsymbol{{\bf \xi}}_{i}.\label{eq:overdamped}
\end{equation}

(\emph{ii}) \emph{linearization}: in the next step, the equations
are linearized in the small displacements $\boldsymbol{r}_{i}(t)=\boldsymbol{r}_{i}^{0}+\boldsymbol{u}_{i}(t)$,
where $\boldsymbol{r}_{i}^{0}+\boldsymbol{u}_{i}^{0}(t)$ is the \emph{instantaneous}
position of the trap center. The mean distance between the trap centers,
$\boldsymbol{\rho}=\boldsymbol{r}_{1}^{0}-\boldsymbol{r}_{2}^{0}$,
is independent of time. This yields a linear equation of motion for
the small displacements $\boldsymbol{u}_{i}$, where the friction
tensors are now evaluated at the mean separation between the traps.
The linearized Langevin equations are
\begin{equation}
\boldsymbol{\gamma}_{ij}(\boldsymbol{r}_{1}^{0},\boldsymbol{r}_{2}^{0})\cdot\dot{\boldsymbol{u}}_{j}+k_{i}\boldsymbol{u}_{i}-k_{i}\boldsymbol{u}_{i}^{0}(t)=\boldsymbol{\xi}_{i}\label{eq:linearized-LE}
\end{equation}
Note that these are 6 coupled stochastic ordinary differential equations. 

(\emph{iii}) \emph{decoupling through the use of symmetries}: in this
step, the symmetry of the friction tensors under translation, assuming
all boundaries are remote, is used to express them as
\begin{equation}
\boldsymbol{\gamma}_{ij}=\gamma_{ij}^{\parallel}(\rho)\hat{\mathbf{\boldsymbol{\rho}}}\hat{\mathbf{\boldsymbol{\rho}}}+\gamma_{ij}^{\perp}(\rho)(\boldsymbol{I}-\hat{\mathbf{\boldsymbol{\rho}}}\hat{\mathbf{\boldsymbol{\rho}}}),\label{eq:gamma-perp-parallel}
\end{equation}
where $\gamma_{ij}^{||}(\rho)$ is the friction coefficient for relative
motion along $\boldsymbol{\rho}$, the line joining the trap centers,
while $\gamma_{ij}^{\perp}(\rho)$ is the corresponding quantity for
motion perpendicular to $\boldsymbol{\rho}$. This motivates the decomposition
of the displacement into components parallel and perpendicular to
$\boldsymbol{\rho},$
\begin{equation}
\boldsymbol{u}_{i}=u_{i}^{\parallel}\hat{\mathbf{\boldsymbol{\rho}}}+\boldsymbol{u}_{i}^{\perp}\cdot(\boldsymbol{I}-\hat{\mathbf{\boldsymbol{\rho}}}\hat{\mathbf{\boldsymbol{\rho}}}).
\end{equation}
Defining the force due to the driving of the trap as $\boldsymbol{f}_{i}(t)=k_{i}\boldsymbol{u}_{i}^{0}(t)$,
averaging the equations over the noise, and using the two previous
equations, we obtain 3 decoupled \emph{pairs} of equations for each
component of motion. For motion along the trap, the pair of coupled
equations is
\begin{equation}
\gamma_{ij}^{||}\dot{u}_{j}^{||}+k_{i}u_{i}^{||}=f_{i}^{||}(t),
\end{equation}
where the dependence of the friction coefficients on relative separation
has been suppressed. The decoupling can be done before the linearization
to give the same result; the two operations commute. 

\emph{(iv}) \emph{response function}: in the final step the coupled
equations are written as
\begin{equation}
\dot{u}_{i}^{||}+A_{ij}u_{j}^{||}=\mu_{ik}^{||}f_{k},
\end{equation}
where $A_{ij}=\mu_{ij}^{||}k_{j}$ is a ``response'' matrix and
the mobility matrix $\mu_{ij}^{||}$ is the inverse of the friction
matrix, $\text{\ensuremath{\gamma_{ik}^{||}\mu_{kj}^{||}}=\ensuremath{\delta_{ij}}}$.
The response function in the frequency domain, then, is \cite{chaikin2000principles}
\begin{equation}
\chi_{ij}^{||}(\omega)=(-i\omega\delta_{ik}+A_{ik})^{-1}\mu_{kj}^{||}.\label{eq:response-function}
\end{equation}
Computing the inverse gives the following expression for the imaginary
part of the response:\begin{widetext}
\begin{align}
\text{Im}\left[\chi_{ij}^{||}(\omega)\right]= & \frac{\omega}{(\text{det}A-\omega^{2})^{2}+(\omega\begin{tabular}{c}
 tr\end{tabular}A)^{2}}\left(\begin{array}{cc}
\begin{tabular}{c}
 \ensuremath{\tfrac{k_{2}}{k_{1}}\mu_{22}^{\parallel}}\end{tabular}\text{det}A+\mu_{11}^{\parallel}\omega^{2} & -\mu_{12}^{\parallel}(\text{det}A-\omega^{2})\\
\, & \,\\
-\mu_{21}^{\parallel}(\text{det}A-\omega^{2}) & \begin{tabular}{c}
 \ensuremath{\tfrac{k_{1}}{k_{2}}\mu_{11}^{\parallel}}\end{tabular}\text{det}A+\mu_{22}\omega^{2}
\end{array}\right).
\end{align}
\end{widetext}The modulus of the response of the first bead to the
driving of the second bead is
\begin{eqnarray}
|\chi_{21}^{||}| & =\Big| & \frac{i\omega\mu_{21}^{\parallel}}{\text{det}A-\omega^{2}-i\omega\begin{tabular}{c}
 tr\end{tabular}A}\Big|,
\end{eqnarray}
which in non-zero only if there is viscous coupling, $\mu_{12}^{\parallel}\neq0$.
The modulus has a maximum at
\begin{equation}
\omega_{res}=\sqrt{\text{det}A}=\sqrt{\mu_{11}^{\parallel}\mu_{22}^{\parallel}k_{1}k_{2}\left(1-\frac{\mu_{12}^{\parallel}\mu_{21}^{\parallel}}{\mu_{11}^{\parallel}\mu_{22}^{\parallel}}\right)}.
\end{equation}

\emph{(v}) \emph{correlation function}: To calculate the correlation
function we set the modulation, $\boldsymbol{u}_{i}^{0}(t)$, of the
traps to zero in Eq.(\ref{eq:linearized-LE}) and project, as before,
to obtain the Langevin equation for parallel displacement fluctuations
\begin{gather}
\boldsymbol{\gamma}_{ij}^{\parallel}\dot{u}_{j}^{\parallel}(t)+k_{i}u_{i}^{\parallel}(t)=\xi_{i}^{\parallel}(t),\label{eq:overdamped-1}
\end{gather}
\begin{gather}
\langle\xi_{i}^{\parallel}(t)\xi_{j}^{\parallel}(t')\rangle=2k_{B}T\gamma_{ij}^{\parallel}\delta(t-t').
\end{gather}
The Fourier amplitudes of the displacements are
\begin{align}
u_{i}^{\parallel}(\omega) & =(-i\omega\delta_{il}+A_{il})^{-1}\mu_{lk}^{||}\xi_{k}^{\parallel}(\omega),\label{eq:fourier-modes}
\end{align}
and the correlation function is then
\begin{gather}
C_{ij}(\omega)=\langle u_{i}^{\parallel}(\omega)u_{j}^{^{\dagger}\parallel}(\omega)\rangle=\nonumber \\
(-i\omega\delta_{il}+A_{il})^{-1}\mu_{lk}^{||}\langle\xi_{k}\xi_{k'}\rangle\mu_{k'm}^{||}(+i\omega\delta_{mj}+A_{mj}^{T})^{-1}.
\end{gather}
Inserting the variance of the noise, the correlation function is\begin{widetext}
\begin{align}
C_{ij}(\omega)= & \frac{2k_{B}T}{(\text{det}A-\omega^{2})^{2}+(\omega\begin{tabular}{c}
 tr\end{tabular}A)^{2}}\left(\begin{array}{cc}
\mu_{22}^{\parallel}k_{2}-i\omega & -\mu_{12}^{\parallel}k_{2}\\
\\
-\mu_{21}^{\parallel}k_{1} & \mu_{11}^{\parallel}k_{1}-i\omega
\end{array}\right)\left(\begin{array}{cc}
\mu_{11}^{\parallel} & \mu_{12}^{\parallel}\\
\\
\mu_{21}^{\parallel} & \mu_{22}^{\parallel}
\end{array}\right)\left(\begin{array}{cc}
\mu_{22}^{\parallel}k_{2}+i\omega & -\mu_{21}^{\parallel}k_{1}\\
\\
-\mu_{12}^{\parallel}k_{2} & \mu_{11}^{\parallel}k_{1}+i\omega
\end{array}\right).
\end{align}
\end{widetext}Completing the matrix multiplications, the final result
is
\begin{align}
C_{ij}(\omega)= & \frac{2k_{B}T}{\omega}\text{Im}\left[\chi_{ij}^{||}(\omega)\right].
\end{align}
This provides an explicit verification of the fluctuation-dissipation
relation for a pair of viscously coupled oscillators \cite{kubo1966fluctuation}. 


\begin{thebibliography}{28}%
\makeatletter
\providecommand \@ifxundefined [1]{%
 \@ifx{#1\undefined}
}%
\providecommand \@ifnum [1]{%
 \ifnum #1\expandafter \@firstoftwo
 \else \expandafter \@secondoftwo
 \fi
}%
\providecommand \@ifx [1]{%
 \ifx #1\expandafter \@firstoftwo
 \else \expandafter \@secondoftwo
 \fi
}%
\providecommand \natexlab [1]{#1}%
\providecommand \enquote  [1]{``#1''}%
\providecommand \bibnamefont  [1]{#1}%
\providecommand \bibfnamefont [1]{#1}%
\providecommand \citenamefont [1]{#1}%
\providecommand \href@noop [0]{\@secondoftwo}%
\providecommand \href [0]{\begingroup \@sanitize@url \@href}%
\providecommand \@href[1]{\@@startlink{#1}\@@href}%
\providecommand \@@href[1]{\endgroup#1\@@endlink}%
\providecommand \@sanitize@url [0]{\catcode `\\12\catcode `\$12\catcode
  `\&12\catcode `\#12\catcode `\^12\catcode `\_12\catcode `\%12\relax}%
\providecommand \@@startlink[1]{}%
\providecommand \@@endlink[0]{}%
\providecommand \url  [0]{\begingroup\@sanitize@url \@url }%
\providecommand \@url [1]{\endgroup\@href {#1}{\urlprefix }}%
\providecommand \urlprefix  [0]{URL }%
\providecommand \Eprint [0]{\href }%
\providecommand \doibase [0]{http://dx.doi.org/}%
\providecommand \selectlanguage [0]{\@gobble}%
\providecommand \bibinfo  [0]{\@secondoftwo}%
\providecommand \bibfield  [0]{\@secondoftwo}%
\providecommand \translation [1]{[#1]}%
\providecommand \BibitemOpen [0]{}%
\providecommand \bibitemStop [0]{}%
\providecommand \bibitemNoStop [0]{.\EOS\space}%
\providecommand \EOS [0]{\spacefactor3000\relax}%
\providecommand \BibitemShut  [1]{\csname bibitem#1\endcsname}%
\let\auto@bib@innerbib\@empty
\bibitem [{\citenamefont {Nyquist}(1928)}]{nyquist1928thermal}%
  \BibitemOpen
  \bibfield  {author} {\bibinfo {author} {\bibfnamefont {H.}~\bibnamefont
  {Nyquist}},\ }\bibfield  {title} {{\bibinfo {title} {Thermal
  agitation of electric charge in conductors},}\ }\href {\doibase
  10.1103/PhysRev.32.110} {\bibfield  {journal} {\bibinfo  {journal} {Phys.
  Rev.}\ }\textbf {\bibinfo {volume} {32}},\ \bibinfo {pages} {110} (\bibinfo
  {year} {1928})}\BibitemShut {NoStop}%
\bibitem [{\citenamefont {Callen}\ and\ \citenamefont
  {Welton}(1951)}]{callen1951}%
  \BibitemOpen
  \bibfield  {author} {\bibinfo {author} {\bibfnamefont {H.~B.}\ \bibnamefont
  {Callen}}\ and\ \bibinfo {author} {\bibfnamefont {T.~A.}\ \bibnamefont
  {Welton}},\ }\bibfield  {title} {{\bibinfo {title} {Irreversibility
  and generalized noise},}\ }\href {\doibase 10.1103/PhysRev.83.34} {\bibfield
  {journal} {\bibinfo  {journal} {Phys. Rev.}\ }\textbf {\bibinfo {volume}
  {83}},\ \bibinfo {pages} {34} (\bibinfo {year} {1951})}\BibitemShut {NoStop}%
\bibitem [{\citenamefont {Kubo}(1966)}]{kubo1966fluctuation}%
  \BibitemOpen
  \bibfield  {author} {\bibinfo {author} {\bibfnamefont {R.}~\bibnamefont
  {Kubo}},\ }\bibfield  {title} {{\bibinfo {title} {The
  fluctuation-dissipation theorem},}\ }\href
  {http://stacks.iop.org/0034-4885/29/i=1/a=306} {\bibfield  {journal}
  {\bibinfo  {journal} {Rep. Prog. Phys.}\ }\textbf {\bibinfo {volume} {29}},\
  \bibinfo {pages} {255} (\bibinfo {year} {1966})}\BibitemShut {NoStop}%
\bibitem [{\citenamefont {Seifert}(2012)}]{seifert2012stochastic}%
  \BibitemOpen
  \bibfield  {author} {\bibinfo {author} {\bibfnamefont {U.}~\bibnamefont
  {Seifert}},\ }\bibfield  {title} {{\bibinfo {title} {Stochastic
  thermodynamics, fluctuation theorems and molecular machines},}\ }\href
  {\doibase 10.1088/0034-4885/75/12/126001} {\bibfield  {journal} {\bibinfo
  {journal} {Rep. Prog. Phys.}\ }\textbf {\bibinfo {volume} {75}},\ \bibinfo
  {pages} {126001} (\bibinfo {year} {2012})}\BibitemShut {NoStop}%
\bibitem [{\citenamefont {Berthier}\ and\ \citenamefont
  {Biroli}(2011)}]{berthier2011theoretical}%
  \BibitemOpen
  \bibfield  {author} {\bibinfo {author} {\bibfnamefont {L.}~\bibnamefont
  {Berthier}}\ and\ \bibinfo {author} {\bibfnamefont {G.}~\bibnamefont
  {Biroli}},\ }\bibfield  {title} {{\bibinfo {title} {Theoretical
  perspective on the glass transition and amorphous materials},}\ }\href
  {\doibase 10.1103/RevModPhys.83.587} {\bibfield  {journal} {\bibinfo
  {journal} {Rev. Mod. Phys.}\ }\textbf {\bibinfo {volume} {83}},\ \bibinfo
  {pages} {587} (\bibinfo {year} {2011})}\BibitemShut {NoStop}%
\bibitem [{\citenamefont {Fodor}\ \emph {et~al.}(2016)\citenamefont {Fodor},
  \citenamefont {Nardini}, \citenamefont {Cates}, \citenamefont {Tailleur},
  \citenamefont {Visco},\ and\ \citenamefont {van Wijland}}]{fodor2016far}%
  \BibitemOpen
  \bibfield  {author} {\bibinfo {author} {\bibfnamefont {{\'E}.}~\bibnamefont
  {Fodor}}, \bibinfo {author} {\bibfnamefont {C.}~\bibnamefont {Nardini}},
  \bibinfo {author} {\bibfnamefont {M.~E.}\ \bibnamefont {Cates}}, \bibinfo
  {author} {\bibfnamefont {J.}~\bibnamefont {Tailleur}}, \bibinfo {author}
  {\bibfnamefont {P.}~\bibnamefont {Visco}}, \ and\ \bibinfo {author}
  {\bibfnamefont {F.}~\bibnamefont {van Wijland}},\ }\bibfield  {title}
  {{\bibinfo {title} {How far from equilibrium is active matter?}}\
  }\href {\doibase 10.1103/PhysRevLett.117.038103} {\bibfield  {journal}
  {\bibinfo  {journal} {Phys. Rev. Lett.}\ }\textbf {\bibinfo {volume} {117}},\
  \bibinfo {pages} {038103} (\bibinfo {year} {2016})}\BibitemShut {NoStop}%
\bibitem [{\citenamefont {Mason}\ and\ \citenamefont {Weitz}(1995)}]{mason95}%
  \BibitemOpen
  \bibfield  {author} {\bibinfo {author} {\bibfnamefont {T.~G.}\ \bibnamefont
  {Mason}}\ and\ \bibinfo {author} {\bibfnamefont {D.~A.}\ \bibnamefont
  {Weitz}},\ }\bibfield  {title} {{\bibinfo {title} {Optical
  measurements of frequency-dependent linear viscoelastic moduli of complex
  fluids},}\ }\href {\doibase 10.1103/PhysRevLett.74.1250} {\bibfield
  {journal} {\bibinfo  {journal} {Phys. Rev. Lett.}\ }\textbf {\bibinfo
  {volume} {74}},\ \bibinfo {pages} {1250--1253} (\bibinfo {year}
  {1995})}\BibitemShut {NoStop}%
\bibitem [{\citenamefont {Levine}\ and\ \citenamefont
  {Lubensky}(2000)}]{levine2000}%
  \BibitemOpen
  \bibfield  {author} {\bibinfo {author} {\bibfnamefont {Alex~J.}\ \bibnamefont
  {Levine}}\ and\ \bibinfo {author} {\bibfnamefont {T.~C.}\ \bibnamefont
  {Lubensky}},\ }\bibfield  {title} {{\bibinfo {title} {One- and
  two-particle microrheology},}\ }\href {\doibase 10.1103/PhysRevLett.85.1774}
  {\bibfield  {journal} {\bibinfo  {journal} {Phys. Rev. Lett.}\ }\textbf
  {\bibinfo {volume} {85}},\ \bibinfo {pages} {1774--1777} (\bibinfo {year}
  {2000})}\BibitemShut {NoStop}%
\bibitem [{\citenamefont {Johnson}(1928)}]{johnson1928thermal}%
  \BibitemOpen
  \bibfield  {author} {\bibinfo {author} {\bibfnamefont {J.~B.}\ \bibnamefont
  {Johnson}},\ }\bibfield  {title} {{\bibinfo {title} {Thermal
  agitation of electricity in conductors},}\ }\href {\doibase
  10.1103/PhysRev.32.97} {\bibfield  {journal} {\bibinfo  {journal} {Phys.
  Rev.}\ }\textbf {\bibinfo {volume} {32}},\ \bibinfo {pages} {97} (\bibinfo
  {year} {1928})}\BibitemShut {NoStop}%
\bibitem [{\citenamefont {Derjaguin}\ and\ \citenamefont
  {Landau}(1941)}]{derjaguin1941theory}%
  \BibitemOpen
  \bibfield  {author} {\bibinfo {author} {\bibfnamefont {B.~V.}\ \bibnamefont
  {Derjaguin}}\ and\ \bibinfo {author} {\bibfnamefont {L.~D.}\ \bibnamefont
  {Landau}},\ }\bibfield  {title} {{\bibinfo {title} {Theory of the
  stability of strongly charged lyophobic sols and the adhesion of strongly
  charged particles in solutions of electrolytes},}\ }\href@noop {} {\bibfield
  {journal} {\bibinfo  {journal} {Acta Physicochim. USSR}\ }\textbf {\bibinfo
  {volume} {14}},\ \bibinfo {pages} {633--662} (\bibinfo {year}
  {1941})}\BibitemShut {NoStop}%
\bibitem [{\citenamefont {Verwey}\ and\ \citenamefont
  {Overbeek}(1948)}]{verwey1948}%
  \BibitemOpen
  \bibfield  {author} {\bibinfo {author} {\bibfnamefont {E.~J.~W.}\
  \bibnamefont {Verwey}}\ and\ \bibinfo {author} {\bibfnamefont {J.~Th.~G.}\
  \bibnamefont {Overbeek}},\ }\href@noop {} {\emph {\bibinfo {title} {Theory of
  the stability of lyophobic colloids}}}\ (\bibinfo  {publisher} {Elsevier,
  Amsterdam},\ \bibinfo {year} {1948})\BibitemShut {NoStop}%
\bibitem [{\citenamefont {Stilgoe}\ \emph {et~al.}(2011)\citenamefont
  {Stilgoe}, \citenamefont {Heckenberg}, \citenamefont {Nieminen},\ and\
  \citenamefont {Rubinsztein-Dunlop}}]{Stilgoe11}%
  \BibitemOpen
  \bibfield  {author} {\bibinfo {author} {\bibfnamefont {A.~B.}\ \bibnamefont
  {Stilgoe}}, \bibinfo {author} {\bibfnamefont {N.~R.}\ \bibnamefont
  {Heckenberg}}, \bibinfo {author} {\bibfnamefont {T.~A.}\ \bibnamefont
  {Nieminen}}, \ and\ \bibinfo {author} {\bibfnamefont {H.}~\bibnamefont
  {Rubinsztein-Dunlop}},\ }\bibfield  {title} {{\bibinfo {title}
  {Phase-transition-like properties of double-beam optical tweezers},}\ }\href
  {\doibase 10.1103/PhysRevLett.107.248101} {\bibfield  {journal} {\bibinfo
  {journal} {Phys. Rev. Lett.}\ }\textbf {\bibinfo {volume} {107}},\ \bibinfo
  {pages} {248101} (\bibinfo {year} {2011})}\BibitemShut {NoStop}%
\bibitem [{\citenamefont {Berg-S{\o}rensen}\ and\ \citenamefont
  {Flyvbjerg}(2004)}]{berg2004power}%
  \BibitemOpen
  \bibfield  {author} {\bibinfo {author} {\bibfnamefont {K.}~\bibnamefont
  {Berg-S{\o}rensen}}\ and\ \bibinfo {author} {\bibfnamefont {H.}~\bibnamefont
  {Flyvbjerg}},\ }\bibfield  {title} {{\bibinfo {title} {Power
  spectrum analysis for optical tweezers},}\ }\href {\doibase
  10.1063/1.1645654} {\bibfield  {journal} {\bibinfo  {journal} {Rev. Sci.
  Inst.}\ }\textbf {\bibinfo {volume} {75}},\ \bibinfo {pages} {594--612}
  (\bibinfo {year} {2004})}\BibitemShut {NoStop}%
\bibitem [{\citenamefont {Meiners}\ and\ \citenamefont
  {Quake}(1999)}]{Meiners99}%
  \BibitemOpen
  \bibfield  {author} {\bibinfo {author} {\bibfnamefont {J.-C.}\ \bibnamefont
  {Meiners}}\ and\ \bibinfo {author} {\bibfnamefont {S.~R.}\ \bibnamefont
  {Quake}},\ }\bibfield  {title} {{\bibinfo {title} {Direct
  measurement of hydrodynamic cross correlations between two particles in an
  external potential},}\ }\href {\doibase 10.1103/PhysRevLett.82.2211}
  {\bibfield  {journal} {\bibinfo  {journal} {Phys. Rev. Lett.}\ }\textbf
  {\bibinfo {volume} {82}},\ \bibinfo {pages} {2211} (\bibinfo {year}
  {1999})}\BibitemShut {NoStop}%
\bibitem [{\citenamefont {Gardiner}(1985)}]{gardiner1985handbook}%
  \BibitemOpen
  \bibfield  {author} {\bibinfo {author} {\bibfnamefont {C.~W.}\ \bibnamefont
  {Gardiner}},\ }\href@noop {} {\emph {\bibinfo {title} {Handbook of stochastic
  methods}}},\ Vol.~\bibinfo {volume} {3}\ (\bibinfo  {publisher} {Springer
  Berlin},\ \bibinfo {year} {1985})\BibitemShut {NoStop}%
\bibitem [{\citenamefont {Ladd}(1988)}]{ladd1988}%
  \BibitemOpen
  \bibfield  {author} {\bibinfo {author} {\bibfnamefont {A.~J.~C.}\
  \bibnamefont {Ladd}},\ }\bibfield  {title} {{\bibinfo {title}
  {Hydrodynamic interactions in a suspension of spherical particles},}\ }\href
  {\doibase 10.1063/1.454658} {\bibfield  {journal} {\bibinfo  {journal} {J.
  Chem. Phys.}\ }\textbf {\bibinfo {volume} {88}},\ \bibinfo {pages}
  {5051--5063} (\bibinfo {year} {1988})}\BibitemShut {NoStop}%
\bibitem [{\citenamefont {Mazur}\ and\ \citenamefont
  {Saarloos}(1982)}]{mazur1982}%
  \BibitemOpen
  \bibfield  {author} {\bibinfo {author} {\bibfnamefont {P.}~\bibnamefont
  {Mazur}}\ and\ \bibinfo {author} {\bibfnamefont {W.~van}\ \bibnamefont
  {Saarloos}},\ }\bibfield  {title} {{\bibinfo {title} {Many-sphere
  hydrodynamic interactions and mobilities in a suspension},}\ }\href {\doibase
  10.1016/0378-4371(82)90127-3} {\bibfield  {journal} {\bibinfo  {journal}
  {Physica A: Stat. Mech. Appl.}\ }\textbf {\bibinfo {volume} {115}},\ \bibinfo
  {pages} {21--57} (\bibinfo {year} {1982})}\BibitemShut {NoStop}%
\bibitem [{\citenamefont {Cichocki}\ \emph {et~al.}(1994)\citenamefont
  {Cichocki}, \citenamefont {Felderhof}, \citenamefont {Hinsen}, \citenamefont
  {Wajnryb},\ and\ \citenamefont {Blawzdziewicz}}]{cichocki1994friction}%
  \BibitemOpen
  \bibfield  {author} {\bibinfo {author} {\bibfnamefont {B.}~\bibnamefont
  {Cichocki}}, \bibinfo {author} {\bibfnamefont {B.~U.}\ \bibnamefont
  {Felderhof}}, \bibinfo {author} {\bibfnamefont {K.}~\bibnamefont {Hinsen}},
  \bibinfo {author} {\bibfnamefont {E.}~\bibnamefont {Wajnryb}}, \ and\
  \bibinfo {author} {\bibfnamefont {J.}~\bibnamefont {Blawzdziewicz}},\
  }\bibfield  {title} {{\bibinfo {title} {Friction and mobility of
  many spheres in {S}tokes flow},}\ }\href {\doibase 10.1063/1.466366}
  {\bibfield  {journal} {\bibinfo  {journal} {J. Chem. Phys.}\ }\textbf
  {\bibinfo {volume} {100}},\ \bibinfo {pages} {3780--3790} (\bibinfo {year}
  {1994})}\BibitemShut {NoStop}%
\bibitem [{\citenamefont {Singh}\ and\ \citenamefont
  {Adhikari}(2016{\natexlab{a}})}]{singh2016crystallization}%
  \BibitemOpen
  \bibfield  {author} {\bibinfo {author} {\bibfnamefont {R.}~\bibnamefont
  {Singh}}\ and\ \bibinfo {author} {\bibfnamefont {R.}~\bibnamefont
  {Adhikari}},\ }\bibfield  {title} {{\bibinfo {title} {Universal
  hydrodynamic mechanisms for crystallization in active colloidal
  suspensions},}\ }\href {\doibase 10.1103/PhysRevLett.117.228002} {\bibfield
  {journal} {\bibinfo  {journal} {Phys. Rev. Lett.}\ }\textbf {\bibinfo
  {volume} {117}},\ \bibinfo {pages} {228002} (\bibinfo {year}
  {2016}{\natexlab{a}})}\BibitemShut {NoStop}%
\bibitem [{\citenamefont {Singh}\ and\ \citenamefont
  {Adhikari}(2016{\natexlab{b}})}]{singh2016traction}%
  \BibitemOpen
  \bibfield  {author} {\bibinfo {author} {\bibfnamefont {R.}~\bibnamefont
  {Singh}}\ and\ \bibinfo {author} {\bibfnamefont {R.}~\bibnamefont
  {Adhikari}},\ }\bibfield  {title} {{\bibinfo {title} {Generalized
  {S}tokes laws for active colloids and their applications},}\ }\href
  {http://arxiv.org/abs/1603.05735} {\bibfield  {journal} {\bibinfo  {journal}
  {arXiv:1603.05735}\ } (\bibinfo {year} {2016}{\natexlab{b}})}\BibitemShut
  {NoStop}%
\bibitem [{\citenamefont {Pozrikidis}(1992)}]{pozrikidis1992}%
  \BibitemOpen
  \bibfield  {author} {\bibinfo {author} {\bibfnamefont {C.}~\bibnamefont
  {Pozrikidis}},\ }\href@noop {} {\emph {\bibinfo {title} {Boundary Integral
  and Singularity Methods for Linearized Viscous Flow}}}\ (\bibinfo
  {publisher} {Cambridge University Press},\ \bibinfo {year}
  {1992})\BibitemShut {NoStop}%
\bibitem [{\citenamefont {Kim}\ and\ \citenamefont {Karrila}(1992)}]{kim2005}%
  \BibitemOpen
  \bibfield  {author} {\bibinfo {author} {\bibfnamefont {S.}~\bibnamefont
  {Kim}}\ and\ \bibinfo {author} {\bibfnamefont {S.~J.}\ \bibnamefont
  {Karrila}},\ }\href@noop {} {\emph {\bibinfo {title} {Microhydrodynamics:
  Principles and Selected Applications}}}\ (\bibinfo  {publisher}
  {Butterworth-Heinemann},\ \bibinfo {year} {1992})\BibitemShut {NoStop}%
\bibitem [{\citenamefont {Gardiner}(1984)}]{gardiner1984adiabatic}%
  \BibitemOpen
  \bibfield  {author} {\bibinfo {author} {\bibfnamefont {C.~W.}\ \bibnamefont
  {Gardiner}},\ }\bibfield  {title} {{\bibinfo {title} {Adiabatic
  elimination in stochastic systems. i. formulation of methods and application
  to few-variable systems},}\ }\href {\doibase 10.1103/PhysRevA.29.2814}
  {\bibfield  {journal} {\bibinfo  {journal} {Phys. Rev. A}\ }\textbf {\bibinfo
  {volume} {29}},\ \bibinfo {pages} {2814--2822} (\bibinfo {year}
  {1984})}\BibitemShut {NoStop}%
\bibitem [{\citenamefont {van Kampen}(1981)}]{van1981ito}%
  \BibitemOpen
  \bibfield  {author} {\bibinfo {author} {\bibfnamefont {N.~G.}\ \bibnamefont
  {van Kampen}},\ }\bibfield  {title} {{\bibinfo {title} {It{\^o}
  versus {S}tratonovich},}\ }\href {\doibase 10.1007/BF01007642} {\bibfield
  {journal} {\bibinfo  {journal} {J. Stat. Phys.}\ }\textbf {\bibinfo {volume}
  {24}},\ \bibinfo {pages} {175--187} (\bibinfo {year} {1981})}\BibitemShut
  {NoStop}%
\bibitem [{\citenamefont {van Kampen}(1992)}]{van1992stochastic}%
  \BibitemOpen
  \bibfield  {author} {\bibinfo {author} {\bibfnamefont {N.~G.}\ \bibnamefont
  {van Kampen}},\ }\href@noop {} {\emph {\bibinfo {title} {Stochastic processes
  in physics and chemistry}}},\ Vol.~\bibinfo {volume} {1}\ (\bibinfo
  {publisher} {Elsevier},\ \bibinfo {year} {1992})\BibitemShut {NoStop}%
\bibitem [{\citenamefont {Klimontovich}(1990)}]{klimontovich1990ito}%
  \BibitemOpen
  \bibfield  {author} {\bibinfo {author} {\bibfnamefont {Y.~L.}\ \bibnamefont
  {Klimontovich}},\ }\bibfield  {title} {{\bibinfo {title} {It{\^o},
  {S}tratonovich and kinetic forms of stochastic equations},}\ }\href {\doibase
  10.1016/0378-4371(90)90142-F} {\bibfield  {journal} {\bibinfo  {journal}
  {Physica A: Stat. Mech. Appl.}\ }\textbf {\bibinfo {volume} {163}},\ \bibinfo
  {pages} {515--532} (\bibinfo {year} {1990})}\BibitemShut {NoStop}%
\bibitem [{\citenamefont {Klimontovich}(1994)}]{klimontovich1994nonlinear}%
  \BibitemOpen
  \bibfield  {author} {\bibinfo {author} {\bibfnamefont {Y.~L.}\ \bibnamefont
  {Klimontovich}},\ }\bibfield  {title} {{\bibinfo {title} {Nonlinear
  {B}rownian motion},}\ }\href {\doibase 10.1070/PU1994v037n08ABEH000038}
  {\bibfield  {journal} {\bibinfo  {journal} {Physics-Uspekhi}\ }\textbf
  {\bibinfo {volume} {37}},\ \bibinfo {pages} {737--766} (\bibinfo {year}
  {1994})}\BibitemShut {NoStop}%
\bibitem [{\citenamefont {Chaikin}\ and\ \citenamefont
  {Lubensky}(2000)}]{chaikin2000principles}%
  \BibitemOpen
  \bibfield  {author} {\bibinfo {author} {\bibfnamefont {P.~M.}\ \bibnamefont
  {Chaikin}}\ and\ \bibinfo {author} {\bibfnamefont {T.~C.}\ \bibnamefont
  {Lubensky}},\ }\href@noop {} {\emph {\bibinfo {title} {Principles of
  condensed matter physics}}},\ Vol.~\bibinfo {volume} {1}\ (\bibinfo
  {publisher} {Cambridge Univ Press},\ \bibinfo {year} {2000})\BibitemShut
  {NoStop}%
\end{thebibliography}
\end{document}